\documentclass[prd,aps,twocolumn,showpacs,preprintnumbers,amsmath,amssymb,nofootinbib]{revtex4}
\usepackage[dvips]{graphicx}
\usepackage{epsf}
\usepackage{amsmath}
\usepackage{amssymb}
\usepackage{tensor}
\usepackage{graphicx}% Include figure files
\usepackage{dcolumn}% Align table columns on decimal point
\usepackage{bm}% bold math
\usepackage{color}
\usepackage[colorlinks]{hyperref}
\begin{document}
\title{Testing  $f(R)$ dark energy model with the large scale structure}

\author{Jian-hua He$^{1}$\email{jianhua.he@brera.inaf.it}}

\affiliation{$^{1}$ INAF-Osservatorio Astronomico
di Brera, Via Emilio Bianchi, 46, I-23807, Merate
(LC), Italy}


\begin{abstract}
In this work, we further investigate the family of $f(R)$ dark energy models that can exactly mimic the same background expansion history as that of the $\Lambda$CDM model. We study the large scale structure in the $f(R)$ gravity using the full set cosmological perturbation equations. We investigate the structure formation in both the spatially flat and curved Universe. We also confront our model with the latest observations and conduct a Markov Chain Monte Carlo analysis on the parameter space.
\end{abstract}

\maketitle

\section{Introduction}
There has been accumulated conclusive evidence from supernovae\cite{1} and other observations\cite{WMAP,BAOm} in the last decade, indicating that our Universe is undergoing a phase of accelerated expansion. Understanding the nature of the cosmic acceleration is one of the biggest questions in modern physics.

The leading explanation of the accelerated expansion is the cosmological constant within the context of General Relativity.  However, the measured value of cosmological constant is far below the prediction of any sensible quantum field theories and the cosmological constant will inevitably lead to the coincidence problem that why the energy density of matter and vacuum are in the same order today(see \cite{sean} for review). Another possibility for the acceleration is that the universe is driven by a new and yet-unknown component called dark energy. The dark energy is some kind of dynamical fluid with negative and time-dependent equation of state $w(a)$. However, the nature of the dynamical dark energy is even harder to be understood in the fundamental physics than that of the cosmological constant.

Alternatively, a promising explanation of the acceleration is the modified gravity. The General relativity might not be ultimately correct in the cosmological scale. The Universe might be described by some kind of modified gravity.
One simplest attempt is called $f(R)$ gravity, in which the
scalar curvature in the Lagrange density of
Einstein's gravity is replaced by an arbitrary
function of $R$\cite{fr}. The $f(R)$ gravity can produce the accelerated expansion of the Universe with any designed effective dark energy equation of state $w$\cite{Song}. Furthermore, the time dependent effective DE EoS in the Jordan frame can be reproduced from the dilation of the inertial mass in the Einstein frame through the conformal transformation \cite{dicke}\cite{Fujii}\cite{Hefr}. In this sense, the time dependent dark energy phenomenon can be better understood in a physical way in the framework of modified gravity.

In this paper, we investigate a specific family of $f(R)$  models that can reproduce the same background expansion history as that of the $\Lambda$CDM model since it has been argued that a valid $f(R)$ model should closely math the $\Lambda$CDM background\cite{Husawicki}. The family of $f(R)$ model contains only one more extra parameter than that of the $\Lambda$CDM model and furthermore, it does have the well-defined Lagrangian formalism in the spatially flat Universe, which is valid for the whole expansion history of the Universe from the past to the future. The model is no longer simply a phenomenological model. Although this model has been studied by a number of work within the parameterized framework of modified gravity~\cite{PPF}~\cite{gbz}, it still needs to solve the full set of cosmological perturbation equations to get more accurate results on the scale $k$ dependent growth history of the Universe when  confronted with upcoming high precision data in cosmological surveys. Therefore, in this work, instead of using the parameterized framework of modified gravity, we investigate the impact of $f(R)$ models on the large scale structure using the full set of covariant cosmological perturbation equations.  We will confront our $f(R)$ model with the latest observations and conduct a Markov Chain Monte Carlo analysis on the parameter space.  We will also exploit the spatially non-flat case in the $f(R)$ gravity, which has not been addressed in the previous work~\cite{gbz}~\cite{songfitting}~\cite{Lucas}.

This paper is organized as follows: In section~\ref{background}, we review the background dynamics of the Universe in the $f(R)$ gravity and present the well-defined Lagrangian formalism in the spatially flat Universe for the  family of $f(R)$ models that reproduce the $\Lambda$CDM background expansion history.
In section~\ref{perturbations}, we present the scalar perturbation equations in the synchronous gauge for the $f(R)$ gravity and study the impact of $f(R)$ gravity on the large scale structure formation.
In section~\ref{constrains}, we present the fittings results by confronting our model with the latest observations.  In section~\ref{conclusions}, we summarize and conclude this work.
\section{The background dynamics\label{background}}
We work with the 4-dimensional action in the $f(R)$ gravity\cite{frreview}
\begin{equation}
S=\frac{1}{2\kappa^2}\int d^4x\sqrt{-g}f(R)+\int d^4x\mathcal{L}^{(m)}\quad,
\end{equation}
where $\kappa^2=8\pi G$.  We consider a homogeneous and isotropic background universe
described by the
Friedmann-Robertson- Walker(FRW) metric
\begin{equation}
ds^2=a^2[-d\tau^2+d\sigma^2]\quad,
\end{equation}
where $d\sigma^2$ is the conformal space-like hypersurface with a constant curvature $R^{(3)}=6K$
\begin{equation}
d\sigma^2=\frac{dr^2}{1-Kr^2}+r^2(d\theta^2+sin^2\theta d\phi^2)\quad .
\end{equation}
The dynamics of the Universe in the $f(R)$ gravity is described by\cite{frreview}
\begin{equation}
\ddot{F}+2F\dot{H}-H\dot{F}-\frac{2K}{a^2}F=-\kappa^2(\rho+p)\quad.\label{dfield}
\end{equation}
where $F=\frac{df(R)}{dR}$, the dot denotes the time derivative with respect to the cosmic time $t$ and $\rho $ is the total energy density of the matter which consists of
the cold dark matter, baryon and radiation
$\rho=\rho_c+\rho_b+\rho_r$. $p$ is the total pressure in the Universe.
If we convert the derivatives in Eq.(\ref{dfield}) from the cosmic time $t$ to $x=\ln a$ ,
Eq.(\ref{dfield}) can be recast into
\begin{equation}
\begin{split}
&\frac{d^2}{dx^2}F+(\frac{1}{2}\frac{d\ln E}{dx}-1)\frac{dF}{dx}+(\frac{d\ln E}{dx}-\frac{2K}{E}e^{-2x})F\\
&=\frac{\kappa^2}{3E}\frac{d\rho}{dx}\quad ,\label{Ffield}
\end{split}
\end{equation}
where
\begin{equation}
\begin{split}
E&\equiv\frac{H^2}{H_0^2}\quad ,\\
R&\equiv3(\frac{d E}{dx}+4E)+6Ke^{-2x}\quad ,\\
\frac{d\rho}{dx}&=-3(\rho+p)\quad.\label{Rdef}
\end{split}
\end{equation}

For convenience, the energy density $\rho$, $K$ and the scalar curvature $R$ in Eq.(\ref{Ffield}) are in the unit of $H_0^2$ and we also set $\kappa^2=1$ in our analysis. In order to get a viable $f(R)$ model with a reasonable expansion history of the Universe and without loss of generality, we can parameterize the quantity $E(x)$ in Eq.(\ref{Ffield}) as the standard model in Einstein's gravity with an effective dark energy equation of state(EoS) $w$\cite{Song}\cite{Pogosian}
\begin{equation}
E(x)=\Omega_r^0e^{-4x}+\Omega_m^0e^{-3x}+\Omega_k^0e^{-2x}+\Omega_d^0e^{-3\int_0^x(1+w)dx}\quad,\label{Ex}
\end{equation}
where
\begin{equation}
\begin{split}
\Omega_k^0&\equiv-\frac{K}{H_0^2}\quad ,\\
\Omega_m^0&\equiv\frac{\kappa^2\rho_m^0}{3H_0^2}\quad,\\
\Omega_d^0&\equiv\frac{\kappa^2\rho_d^0}{3H_0^2}\quad,\\
\Omega_r^0&\equiv\frac{\kappa^2\rho_r^0}{3H_0^2}\quad.\label{defination}
\end{split}
\end{equation}
$\rho_d^0$ indicates the energy density of the effective dark energy, $\rho_m^0=\rho_b^0+\rho_c^0$ is the energy density of non-relativistic matter and $\rho^0=\rho_m^0+\rho_r^0$ is the total energy density of the matter.

The $f(R)$ model can be constructed by specifying the background expansion history of the Universe in
Eq.(\ref{Ffield}). Eq.(\ref{Ffield}) becomes a second order differential equation only with respect to $F(x)$.
However, any specifically designed time dependent effective dark energy equation of state can hardly be well motivated in physics because we still have less knowledge about the nature of the dark energy at present.
Therefore, it is of great interest to investigate the simplest case that the $f(R)$ model can reproduce the same background expansion history as that of the $\Lambda$CDM model $w=-1$. Therefore, in this work we will only focus on this case hereafter.

In the spatially flat Universe $K=0$ filled with dust matter, the late time expansion history of the Universe with the effective dark energy EoS $w=-1$ can be written as,
\begin{equation}
E(x)=\Omega_m^0e^{-3x}+\Omega_d^0\quad.\label{Ex}
\end{equation}
In this case, Eq.(\ref{Ffield}) can be solved analytically. The general solutions for  Eq.(\ref{Ffield}) are
\begin{equation}
\begin{split}
F(x)-1&= C(e^{3x})^{p_-}{_2F_1}[q_-,p_-;r_-;-e^{3x}\frac{\Omega_d^0}{\Omega_m^0}]\\
&+D(e^{3x})^{p_+}{_2F_1}[q_+,p_+;r_+;-e^{3x}\frac{\Omega_d^0}{\Omega_m^0}] \quad\label{mG},
\end{split}
\end{equation}
where the indexes in the above expressions are given by
\begin{eqnarray}
q_\pm&=&\frac{1\pm\sqrt{73}}{12}\nonumber \quad,\\
r_\pm&=&1\pm\frac{\sqrt{73}}{6}\nonumber \quad,\\
p_\pm&=&\frac{5\pm\sqrt{73}}{12}\nonumber \quad.
\end{eqnarray}
${_2F_1}[a,b;c;z]$ is the Gaussian hypergeometric function. $C$ and $D$ are arbitrary constants.
A viable $f(R)$ model should be in a ``chameleon" type \cite{Mota}\cite{Khoury}
where in the past of the Universe, the model should go back to the standard model
\begin{equation}
\lim_{x\rightarrow-\infty}F(x)=1\quad.\label{boundary}
\end{equation}
The first term in Eq.(\ref{mG}) is divergent when $x$ goes to the minus infinity due to the negative index $p_-$ and
$C$, therefore, should vanish  $C=0$ to satisfy the boundary condition Eq.(\ref{boundary}).

After we get the expression for $F(x)$, we can obtain the explicit expression for $f(R)$ by doing the integration
\begin{equation}
f(R)=\int F(x)\frac{dR}{dx}dx\quad\label{generalfr}\quad,
\end{equation}
and using the relationship between $R$ and $x$.
\begin{equation}
R=3\Omega_m^0e^{-3x}+12\Omega_d^0\quad.\label{Rscalar}
\end{equation}
The final result for $f(R)$ turns out to be
\begin{equation}
\begin{split}
f(R)&=R-2\Lambda\\
&-\varpi\left (\frac{\Lambda}{R-4\Lambda}\right )^{p_+-1}{_2F_1}\left[q_+,p_+-1;r_+;-\frac{\Lambda}{R-4\Lambda}\right ]
\end{split}\quad,\label{viable}
\end{equation}
where $\Lambda$, $\varpi$ are constant parameters.
\begin{equation}
\varpi= D(R_0-4\Lambda)^{p_+}/(p_+-1)/\Lambda^{p_+-1}\quad.
\end{equation}
When $\varpi=0$, $\Lambda $ is just the cosmological constant.
If we write Eq.(\ref{viable}) in the units of $H_0^2$, $R_0$ and $\Lambda$ can be presented as $R_0= (3\Omega_m^0+12\Omega_d^0)$ and $\Lambda=3\Omega_d^0$ respectively.

The hypergeometric function ${_2F_1}[a,b;c;z]$ can have the integral representation on the real axis when $b>0$ and $c>0$
\begin{equation}
{_2F_1}[a,b;c;z]=\frac{\Gamma(c)}{\Gamma(b)\Gamma(c-b)}\int_0^{1}t^{b-1}(1-t)^{c-b-1}(1-zt)^{-a}dt\quad,\label{defhypergeometric}
\end{equation}
where $\Gamma$ is the Euler Gamma function. In this case, ${_2F_1}[a,b;c;z]$ is well-defined in the range $-\infty<z<1$
The expression of Eq.(\ref{viable}), therefore, is a well-defined real function on the real axes in the physical range $R>4\Lambda$, which is different from the results given by~\cite{solution}. A more detailed analysis on Eq.(\ref{viable}) has been presented in our companion work~\cite{frlcdm}.

In the past of the Universe, when $R>>4\Lambda$, the hypergeometric function goes back to unity ${_2F_1}\sim1$,
and Eq.(\ref{viable}) becomes
\begin{equation}
f(R)\sim R-\varpi\left (\frac{\Lambda}{R}\right )^{p_+-1}\quad.
\end{equation}
The above expression can further go back to the standard model $f(R)\sim R$ when $R$ becomes more larger.
On the other hand, in the future expansion, Eq.(\ref{viable}) does not have the future singularity because $f(R)$ is finite at the point of $R=4\Lambda$.
\begin{equation}
\begin{split}
&\lim_{R\rightarrow4\Lambda}f(R)=2\Lambda-\\
&\frac{\varpi4(-511+79\sqrt{73})\Gamma(2/3)\Gamma(-r_-)}{(-5+\sqrt{73})(-1+\sqrt{73})(7+\sqrt{73})\Gamma(-p_-)\Gamma(q_+)}\\
&\approx2\Lambda-1.256\varpi
\end{split}\quad.
\end{equation}
When $R<4\Lambda$ the expression of Eq.(\ref{viable}) becomes complex, which is clearly unphysical.

\begin{figure}
\centering
\includegraphics[width=3.2in,height=2.8in]{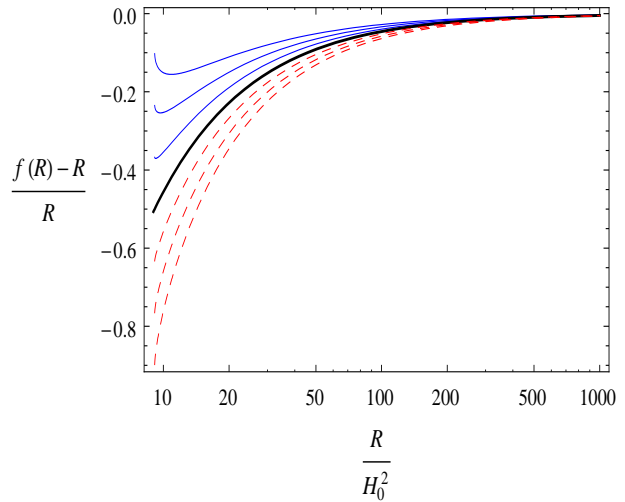}
\caption{The f(R) models that can reproduce the $\Lambda$CDM expansion history.
From top $D=-0.6,-0.4,-0.2,0,0.2,0.4,0.6$ and $\Omega_m^0=0.24$}\label{frmodels}
\end{figure}

In the spatially flat Universe, Eq.(\ref{viable}) is valid for the whole expansion history of the Universe from the radiation dominated epoch to the future expansion of the Universe. Eq.(\ref{viable}) can exactly mimic the $\Lambda$CDM expansion history from the matter dominated epoch to the late time acceleration. For illustrative propose, in Fig.~\ref{frmodels}, we plot the $f(R)$ models for a few representative values of $D$. The curves in Fig.~\ref{frmodels} are evaluated from Eq.(\ref{viable}) directly. By noting the different conventions of $f(R)$ used in our work and \cite{Song}, our results are consistent with the numerical results presented in \cite{Song}.

As shown in~\cite{Ignacy}, in order to avoid the instabilities in the high curvature region, it requires that $F>0$ and $f_{RR}=\frac{\partial F}{\partial R}>0$. Therefore, in our model $D$  should be negative $D<0$. The models presented in the red dashed lines in Fig.~\ref{frmodels}($D>0$) are ruled out by the instabilities in the high curvature region. Since the $f(R)$ model investigated in our work has only one more extra parameter than that of the $\Lambda$CDM model, the family of $f(R)$ models can be characterized by either $D$ or $B_0=\frac{f_{RR}}{F}\frac{dR}{dx}H/\frac{dH}{dx}(x=0)$\cite{Song}. In the appendix, we provide the explicit relationship between $B_0$ and $D$.

However, in this work, we are not only limited in the spatially flat Universe, we will also  exploit the $f(R)$ model in the spatially curved Universe. Without losing generality, we will numerically find the solution of Eq.(\ref{Ffield}) to include the spatially curved case $\Omega_k^0\neq0$. We perform our numerical calculation starting from the deep matter dominated epoch($a_i\sim0.03$). The curvature term $K$ can be neglected at this epoch and the analytical solutions for Eq.(\ref{Ffield}) give rise to
\begin{equation}
\begin{split}
F(x)&\sim1+D(e^{3x})^{p_+}\quad,\\
\frac{dF(x)}{dx}&\sim3Dp_+(e^{3x})^{p_+}\quad,
\end{split}
\end{equation}
We take the above expressions as the initial conditions for Eq.(\ref{Ffield}).
In this work, we use the parameter $D$ to characterize the family of $f(R)$ models and treat $B_0$ as derived parameter since $D$ directly relates to the covariant parameter $\varpi$ for this kind of $f(R)$ models in the spatially flat Universe. Since the $f(R)$ model investigated in our work and the $\Lambda$CDM model can only be distinguished in their perturbed space-time, in the next section, we will turn to the cosmological perturbations theory.
\section{cosmological perturbations\label{perturbations}}
The cosmological perturbation theory for the $f(R)$ gravity has been well studied in~\cite{frperturbationreview}.  ~\cite{frperturbationreview} has extensively presented the perturbation equations for a wide family of modified gravities.  The perturbation equations in the spatially flat Universe for the $f(R)$ gravity can be found in~\cite{bean}.

For the scalar perturbation, the perturbed line element can be written as\cite{Kodama}
\begin{eqnarray}
ds^2 &=&  a^2[-(1+2\psi Y^{(s)})d\tau^2+2BY^{(s)}_id\tau
dx^i\nonumber \\
&+&(1+2\phi Y^{(s)})\gamma_{ij}dx^idx^j+\mathcal{E}Y^{(s)}_{ij}dx^idx^j]\label{pert_jordan}\quad,
\end{eqnarray}
where $\gamma_{ij}$ in the spherical coordinate can be written as
\begin{equation}
[\gamma_{ij}]=\begin{pmatrix} \frac{1}{1-Kr^2} & 0 & 0 \\
               0 & r^2 & 0 \\
               0 & 0 & r^2sin^2\theta
\end{pmatrix}\quad,
\end{equation}
$Y$,$Y_{j}$ and $Y_{ij}$  are the scalar harmonic functions which are defined by
\begin{equation}
\begin{split}
&(\Delta+k^2)Y^{(s)}=0\quad ,\\
&Y_j^{(s)}\equiv-\frac{1}{k}Y_{|j}^{(s)}\quad ,\\
&Y_{ij}^{(s)}\equiv\frac{1}{k^2}Y_{|ij}^{(s)}+\frac{1}{3}\gamma_{ij}Y^{(s)}\quad.
\end{split}
\end{equation}
The detailed general covariant perturbation equations including the spatially curved case for the $f(R)$ gravity are presented in the appendix.

In this work, we focus on the synchronous gauge which is defined by $\psi=0$ and $B=0$. The synchronous gauge is widely used in the Einstein-Boltzmann codes~\cite{cmbfast}~\cite{easy}~\cite{CAMB}~\cite{Class} to calculate the temperature and polarization power spectra of the cosmic microwave background anisotropy.

\begin{figure}
\includegraphics[width=3.2in,height=2.8in]{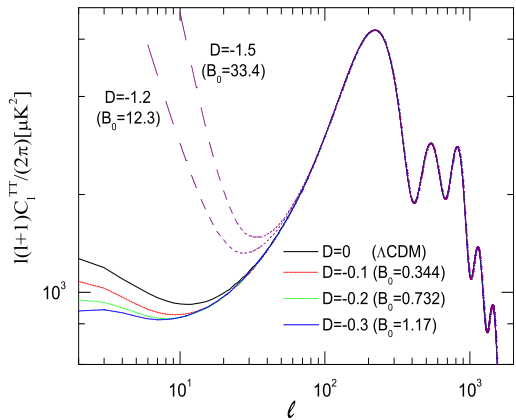}
\caption{The angular power spectrum of the temperature correlation
}\label{angular}
\end{figure}

The synchronous gauge is characterized by the following two parameters
\begin{eqnarray}
\eta_T&=&-(\phi+\frac{\mathcal{E}}{6})\nonumber\quad,\\
h_L&=&6\phi\quad,
\end{eqnarray}
where $\eta_T$ refers to the conformal 3-space curvature perturbation
\begin{equation}
\delta R^{(3)}=6\delta K=-4(k^2-3K)\eta_T\quad.
\end{equation}
The perturbed modified Einstein equations in the synchronous can be written as
\begin{eqnarray}
-\frac{1}{2}\kappa^2a^2\delta \rho&=&-(\frac{1}{2}F\mathcal{H}+\frac{1}{4}F')h_L'-\frac{3}{2}\mathcal{H}\delta F'-\frac{1}{2}\delta F k^2\nonumber \\
&+& F \eta_T(k^2-3K)+\frac{3}{2}\mathcal{H}'\delta F\label{SynE1}\quad, \\
\kappa^2a^2\delta p&=&F[-\frac{2}{3}\mathcal{H}h_L'+\frac{2}{3}k^2\eta_T-\frac{1}{3}h_L''-2\eta_TK]\nonumber \\
&+&\delta F[\mathcal{H}^2+\frac{a''}{a}-\frac{2}{3}k^2+2K]-\frac{1}{3}F'h_L'\nonumber \\
&-&\delta F''-\delta F'\mathcal{H}\quad,\label{SynE2}\\
\alpha'&=&-2\mathcal{H}\alpha+\eta_T-\frac{F'}{F}\alpha -\kappa^2a^2\frac{p\Pi}{Fk^2}-\frac{\delta F}{F}\nonumber\quad ,\\\label{SynE3}
\\
\frac{k^2-3K}{k}F\eta_T'&=&\frac{1}{2}\kappa^2a^2q+\frac{1}{2}k\delta F'-\frac{1}{2}k\mathcal{H}\delta F+\frac{Fh_L'K}{2k}\nonumber\quad ,\\
\label{SynE4}
\end{eqnarray}
where
\begin{equation}
\begin{split}
\alpha&\equiv\frac{(h_L+6\eta_T)'}{2k^2}\quad,\\
      q&=(\rho+p)v
\end{split}
\end{equation}
and
\begin{eqnarray}
&&\delta F''+2\mathcal{H}\delta F'+a^2(\frac{k^2}{a^2}+M^2)\delta F\nonumber\\
 &=&\frac{\kappa^2a^2}{3}(\delta \rho -3\delta p) -\frac{1}{2}F'h_L'\quad,\label{deltF}
\end{eqnarray}
where
\begin{equation}
M^2=\frac{1}{3}(\frac{F}{f_{RR}}-R)\label{Mdef}\quad.
\end{equation}
The scalar curvature perturbation $\delta R$ is given by
\begin{equation}
a^2\delta R=h_T''+3\mathcal{H}h_T'-4\eta_Tk^2+12K\eta_T\quad.
\end{equation}
With the perturbation equations, we perform our numerical analysis based on the public available Einstein-Boltzmann code CAMB ~\cite{CAMB}.
In the appendix, we summarize the details on how to modify the code.
In Fig.\ref{angular}, we show the full spectrum of temperature anisotropy $C_l^{TT}$ for a few representative values of $D$. The cosmological parameters used in the numerical process are taken from the WMAP 7-year best-fitted values for the $\Lambda$CDM model \cite{WMAP} $\Omega_bh^2=0.0227,\Omega_ch^2=0.112,n_s=0.966,\Delta^2_{\mathcal{R}}=2.42\times10^{-9},\Omega_{\Lambda}=0.729,\tau=0.085$
. From Fig.\ref{angular}, we can see that the $f(R)$ gravity only affects the lowest multipoles in the power spectrum while not change the acoustic peaks. This observation is consistent with what found in\cite{Song}.  The small $l$ power is suppressed as the decreasing of the value of $D$. However, there is a turning around point, roughly about $D<-0.37(B_0>1.5)$, after which there is a prominent enhancement in the power as further decreasing the value of $D$. Numerical results of our work agree with the already presented results within the parameterized framework of modified gravity\cite{PPF}\cite{gbz}.

\begin{figure}
\includegraphics[width=3.2in,height=2.8in]{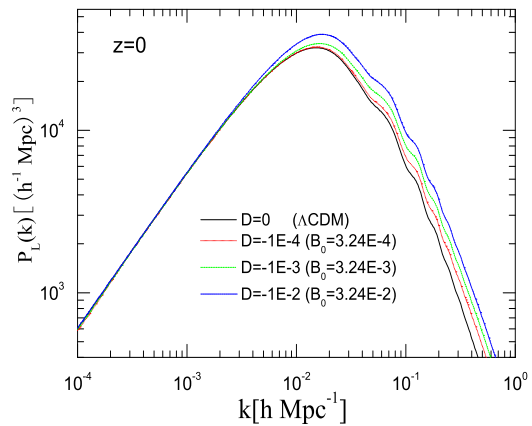}
\caption{The linear matter power spectrum.
}\label{matter}
\end{figure}

\begin{figure}
\includegraphics[width=3.2in,height=2.8in]{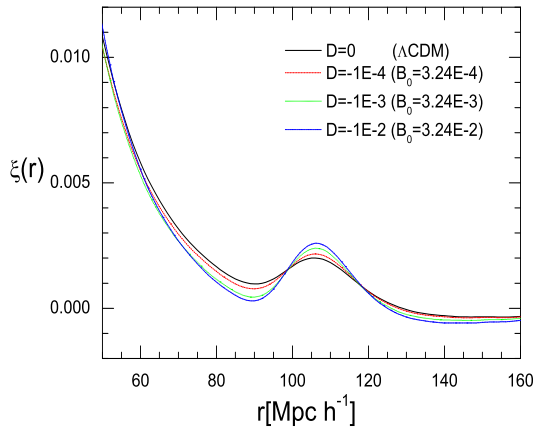}
\caption{The two-point correlation function in the real space
}\label{correlation}
\end{figure}

In Fig.\ref{matter}, we show the linear matter power spectrum at redshift zero $z=0$. The $f(R)$ gravity changes not only the amplitude but also the shape of the power spectrum due to the scale $k$ dependent growth history \cite{Song}. However, the $f(R)$ model investigated in our work should not change the position of the Baryon Acoustic Oscillation(BAO) peak in the two-point correlation function in the real space because the sound horizon is only determined by the background cosmological parameters\cite{BAOreview}\cite{BAO} and the family of $f(R)$ models investigated in our work has the same background expansion.
By doing the Fourier transformation
\begin{equation}
\xi(r)=\frac{1}{2\pi^2}\int dk k^2P_L(k)\frac{\sin(kr)}{kr}\quad,
\end{equation}
in Fig~\ref{correlation}, we show the two-point correlation function in the real space. We can see clearly that
although the shape of the correlation function has changed a lot for a few representative values of $D$, the position of the  BAO peak does not change under the Fourier transformation, which shows that our numerical results are consistent with the theoretical prediction.

\begin{figure}
\includegraphics[width=3.2in,height=2.8in]{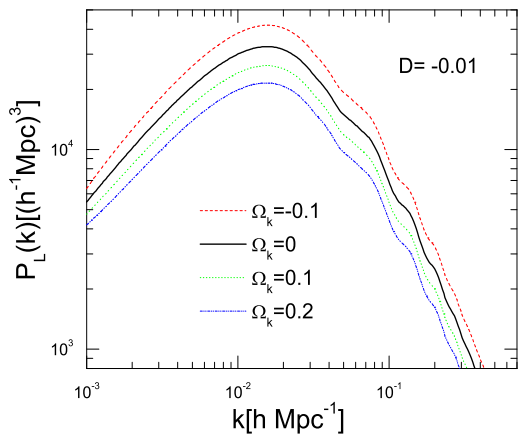}
\caption{The impact of the curvature $\Omega_k$ on the matter power spectrum
}\label{curture}
\end{figure}

In the spatially curved Universe, $\Omega_k^0$ has significant impact on both the matte power spectrum and the CMB angular power spectrum. In the matter power spectrum, as shown in Fig.\ref{curture}, the positive $\Omega_k^0>0$ will suppress the power at all scales and the negative $\Omega_k^0<0$ will oppositely enhance the power. The slight positive value of $\Omega_k^0>0$ could compensate part of the impact induced by the modified gravity $D$ on the matte power spectrum, which enhance the power at scale $k>0.001{\rm h Mpc^{-1}}$(see Fig.\ref{matter}).
Meanwhile, the spatial curvature $\Omega_k^0$ also shifts the positions of acoustic peaks in the CMB temperature angular power spectrum. However, the high precision measurement of the acoustic peaks in the CMB temperature angular power spectrum in combination with BAO and $H_0$\cite{WMAP} can put very tight constrains on the spatial curvature. The spatial curvature, therefore, could not affect significantly on the large scale structure in the $f(R)$ gravity.

\begin{figure}
\includegraphics[width=3.2in,height=2.8in]{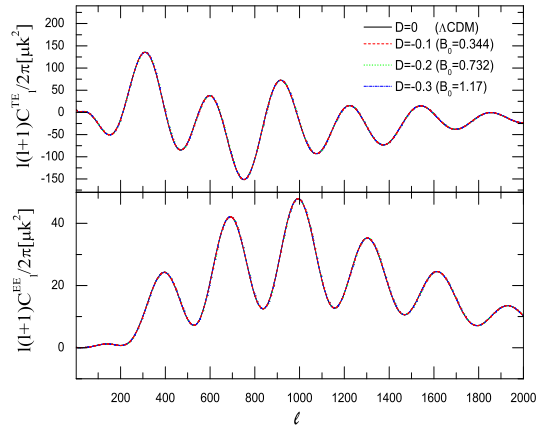}
\caption{The TE and EE angular power spectrum.
}\label{EE}
\end{figure}

Compared with the CMB temperature angular power spectrum $C_l^{TT}$(see Fig.\ref{angular}), there are no significant imprint of $f(R)$ gravity on the temperature polarization cross-correlation and polarization auto-correlation spectrum $C_l^{TE}$,$C_l^{EE}$(see Fig.\ref{EE}) because the CMB polarization anisotropy only arises from the quadrupole anisotropy at the last scattering surface and does not correlate with the late time ISW effect. There are no gravitational perturbation terms appearing in the source term $S_E$ for the polarization correlation. For instance, in the spatially flat Universe, $S_E$ can be written as \cite{source}\cite{cambequations}
\begin{equation}
S_E=\frac{g\zeta}{4k^2(\tau_0-\tau)^2}\quad,
\end{equation}
where $g$ is the visibility function and $\zeta$ is given by
$\zeta=(\frac{3}{4}I_2 +\frac{9}{2}E_2)$. $I_2$, $E_2$ indicate the quadrupole of the photon intensity and the E-like polarization respectively~\cite{cambequations}. After the last scattering, the visibility function $g\simeq\delta(\tau-\tau_{dec})$ drops almost to be zero. The perturbation of the $f(R)$ gravity has no impact on $S_E$ unless the $f(R)$ model changes the background expansion.

After the qualitative analysis of the impact of  $f(R)$ model on the large scale structure, in the next section, we will present cosmological constrains on the $f(R)$ model from the latest observations.
\section{constrains from observations ~\label{constrains}}
\begin{figure*}
\centering
\includegraphics[width=6in,height=5in]{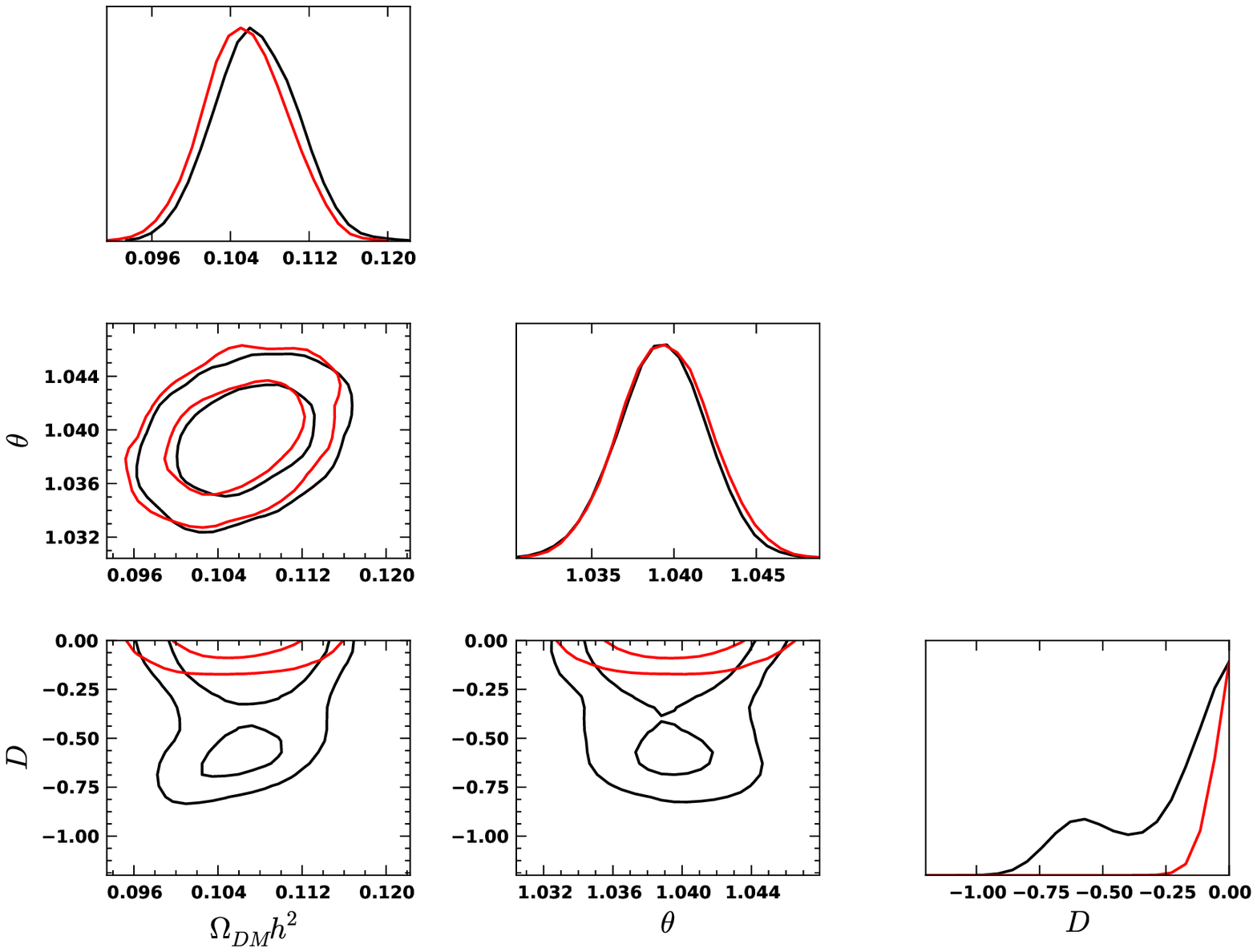}
\caption{Marginalized posterior distribution and 2-D contour plots for $f(R)$ parameters in the flat Universe $\Omega_k=0$. CMB+SN+HST+MPK(black lines), CMB+SN+HST+MPK+gISW(red lines)
}\label{fittingflat}
\end{figure*}

\begin{figure*}
\centering
\includegraphics[width=6in,height=5in]{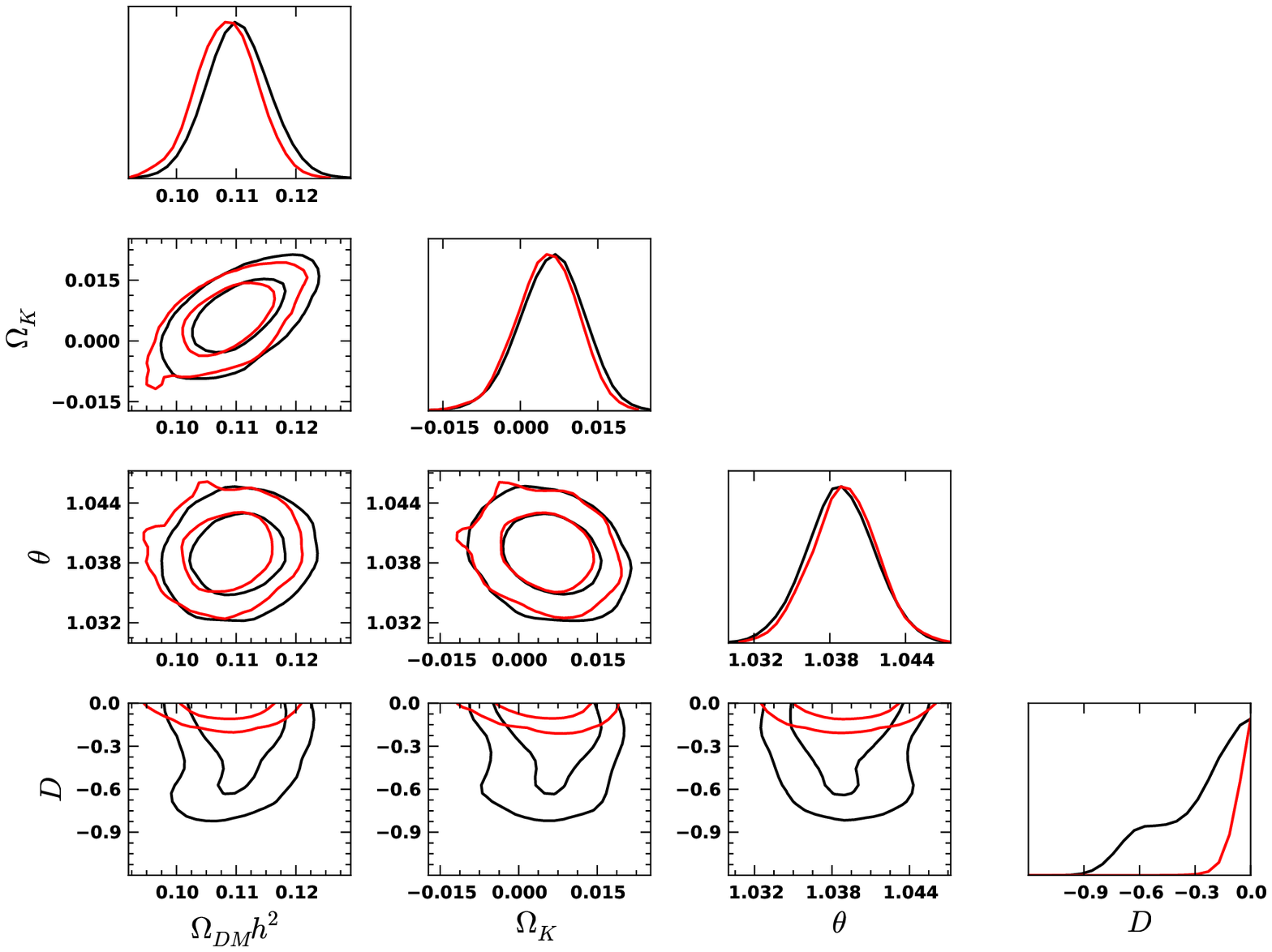}
\caption{Marginalized posterior distribution and 2-D contour plots for $f(R)$ parameters in the non-flat Universe $\Omega_k\neq0$. CMB+SN+HST+MPK(black lines), CMB+SN+HST+MPK+gISW(red lines)
}\label{fittingcur}
\end{figure*}

The parameter space of our model is
\begin{equation}
P=(\Omega_bh^2,\Omega_ch^2,\theta_A,\Omega_k,\ln[10^{10}A_s],n_s,\tau,D)\quad,
\end{equation}
where $\theta_A$ is the angular size of the acoustic horizon, and $A_s$ is amplitude of the primordial curvature perturbation. The priors for cosmological parameters are listed in table~\ref{priors}.
\begin{table*}
\caption{Priors for cosmological parameters \label{priors}}
\begin{tabular}{c}
\hline
$0.005<\Omega_bh^2<0.1$ \\
$0.01<\Omega_ch^2<0.99$\\
$0.5<\theta<10$ \\
$-0.2<\Omega_k<0.2$ \\
$0.01<\tau<0.8$  \\
$0.5<n_s<1.5$  \\
$2.7<\rm{ln}[10^{10}As]<4.0$ \\
$-1.2<D<0$ \\
\hline
\end{tabular}
\end{table*}
For the CMB data, we use the seven-year WMAP(WMAP 7) CMB temperature and polarization power spectra~\cite{WMAP}.  We also take the measurement of the power spectrum of LRG from SDSS DR7 \cite{SDSS}. However, we limit our considerations of the matter power spectrum in a relative linear scale and we only take the samples within bins for $k<0.1h/\rm{Mpc}$.  We add the supernovae data ~\cite{sneia} and the present day value of the Hubble constant from the measurement of Hubble Space Telescope(HST) $H_0=74.2\pm3.6 \rm{kms}^{-1}\rm {Mpc}^{-1}$~\cite{HST} in order to put tighter constrains on the background cosmological parameters. The global fitting results using above data sets are listed in Table~\ref{flat} and ~\ref{cur}. In the flat Universe, we find the constrains on the parameter $D$ as $|D|<0.709(95\%{\rm CL})$ and $|D|<0.711(95\%{\rm CL})$ in the curved Universe. These results are equivalent to $B_0<3.86(95\%{\rm CL})$ and $B_0<3.88(95\%{\rm CL})$ if we use the parameter $B_0$. The result is consistent with \cite{songfitting}. Although the $f(R)$ gravity has significant impact on the shape of the matter power spectrum(see Fig.~\ref{matter}), the SDSS LRG matter spectrum can not put very tight constrains on $D$ because there are degeneracies between $D$ and other cosmological parameters\cite{songfitting}. When $D\sim-0.6$($B_0\sim3$), the CMB temperature angular power spectrum $C_l^{TT}$ of the $f(R)$ model goes back similar to that of $\Lambda$CDM model. Therefore, there are two peaks in the likelihood distribution of $D$ which are clearly shown in the black lines in Fig.\ref{fittingflat} and Fig.\ref{fittingcur}. Clearly, the combination of the data set CMB+SN+HST+MPK can not put very tight constrains on the $f(R)$ models. We need to add additional data set.

As pointed out in \cite{songfitting}, the $f(R)$ gravity can produce the anti-correlation in the Galaxy-ISW angular power spectrum. However, the measurement of the Galaxy-ISW correlation favors the positive correlation.
Therefore, the Galaxy-ISW correlation data set can put very tight constrains on the $f(R)$ model. We consider the cross correlation
\begin{equation}
C_l^{gI}=\frac{2}{\pi}\int k^2dk P_m(k,0)W_l^g(k)W_l^I(k)\quad,\label{CgI}
\end{equation}
where the window functions $W_l^I(k)$ and $W_l^g(k)$ are given by
\begin{eqnarray}
W_l^I(k)&=&-T_0\int dz \frac{d}{dz}(\Psi-\Phi)j_l[k\chi(z)]\nonumber \quad,\\
W_l^g(k)&=&\int dz b(z)\Pi(z)D_g(z)j_l[k\chi(z)]\quad,\label{gicross}
\end{eqnarray}
where $\Psi-\Phi$  can be presented  in terms of the quantities in the synchronous gauge as $\Psi-\Phi=\eta_T+\alpha'$, $\chi(z)$ is the looking back time, $j_l(x)$ is the spherical bessel function, $b(z)$ is the bias, $\Pi(z)$ is the normalized selection function, $D_g(z)$ is defined as $D_g(z)=\frac{\delta_m(z)}{\delta_m(0)}$ and $T_0$ is the temperature of the CMB today.

In order to improve the performance in the numerical process, we use the Limber approximation
\begin{equation}
\frac{2}{\pi}\int k^2dk j_l[k\chi]j_l[k\chi']\approx\frac{\delta(\chi-\chi')}{\chi^2}\quad,
\end{equation}
and Eq.(\ref{CgI}) reduces to
\begin{equation}
C_l^{gI}\sim T_0\int dz \frac{b(z)\Pi(z)}{\chi^2}P_{gI}(\frac{l+1/2}{\chi},z)\quad,
\end{equation}
where
\begin{equation}
P_{gI}(k,z)=\frac{2\pi^2}{k^3}P_{\chi}(k)\delta_m(k,z)\frac{d}{d\tau}(\Psi-\Phi)\quad,
\end{equation}
and
\begin{equation}
P_{\chi}(k)=A_s(\frac{k}{k_*})^{n_s-1}\quad.
\end{equation}

In this work, we adopt the Galaxy-ISW correlation data from \cite{ISW} and use the public available ISWWLL code\cite{ISW} to calculate the likelihood. The bias $b(z)$ and the selection function $\Pi(z)$  are provided by the ISWWLL code and will be recomputed for each time according to different cosmological parameters in the Markov chain.  We have turned off the contribution of weak lensing in the original code. The results of the joint likelihood analysis with the data sets from CMB+SN+HST+MPK+gISW are shown in Table~\ref{flat} and ~\ref{cur}. The marginalized $1D$ and $2D$ likelihoods for interested parameters are shown in red lines in Fig.\ref{fittingflat} and Fig.\ref{fittingcur}. The Galaxy-ISW correlation data set has improved significantly the constrains on the parameter of $D$ up to $|D|<0.109(B_0<0.376)(95\%{\rm CL})$ in the flat Universe and $|D|< 0.131(B_0<0.459)(95\%{\rm CL})$ in the curved Universe. The result in the flat Universe are in good agreement with the work done by~\cite{gbz}and~\cite{Lucas}. The best fitted point for the curvature is slightly positive $\Omega_k=0.0063^{+0.0061}_{-0.0061}$ which is different from the results in the non-flat $\Lambda$CDM model $\Omega_k=-0.0023^{+0.0054}_{-0.0056}$\cite{WMAP}.

Although there has been reported recently that more stringent constrains on the $f(R)$ gravity can be obtained by using the data from cluster abundance\cite{cluster}, we will not include this data in this work because we do not have the reliable knowledge about the halo mass function in our $f(R)$ model. The halo mass function used in \cite{cluster} has been tested by a large suite of N-body simulations and shown to be a reasonable fit to Hu-Sawicki model\cite{cluster}\cite{HuI}. However, in our model, we still need to investigate the halo mass function before the data can be used for our global fitting analysis.

\begin{table*}
\caption{Fitting results for the flat model $\Omega_k=0$ \label{flat}}
\begin{tabular}{|c||c|c|}
\hline
Parameters & CMB+SN+HST+MPK& CMB+SN+HST+MPK+gISW\\
\hline
$\Omega_bh^2$ &$0.02257^{+0.00053}_{-0.00053}$ &$0.02257^{+0.00053}_{-0.00053}$ \\
\hline
$\Omega_ch^2$ &$0.1067^{+0.0040}_{-0.0040}$ &$0.1055^{+0.0041}_{-0.0041}$\\
\hline
$\theta$ & $1.0396^{+0.0026}_{-0.0026}$ & $1.0395^{+0.0027}_{-0.0027}$\\
\hline
$\tau$ &$ 0.091^{+0.015}_{-0.015}$ & $0.091^{+0.015}_{-0.015}$ \\
\hline
$n_s$ & $0.966^{+0.013}_{-0.013}$ & $0.968^{+0.013}_{-0.013}$ \\
\hline
$\rm{ln}[10^{10}As]$ & $3.065^{+0.035}_{-0.035}$ &$3.060^{+0.036}_{-0.036}$ \\
\hline
$|D|$ & $ <0.709$(95\%{\rm CL}) & $ <0.109$(95\%{\rm CL})\\
\hline
\end{tabular}
\end{table*}
\begin{table*}
\caption{Fitting results for the non-flat model $\Omega_k\neq0$ \label{cur}}
\begin{tabular}{|c||c|c|}
\hline
Parameters & CMB+SN+HST+MPK & CMB+SN+HST+MPK+gISW\\
\hline
$\Omega_bh^2$ &$0.02240^{+0.00055}_{-0.00055}$ &$0.02246^{+0.00054}_{-0.00054}$ \\
\hline
$\Omega_ch^2$ &$0.1102^{+0.0052}_{-0.0052}$ &$0.1084^{+0.0050}_{-0.0050}$\\
\hline
$\theta$ & $1.0390^{+0.0027}_{-0.0027}$ & $1.0392^{+0.0025}_{-0.0025}$\\
\hline
$\tau$ &$ 0.088^{+0.014}_{-0.014}$ & $0.089^{+0.015}_{-0.015}$ \\
\hline
$\Omega_k$ & $0.0063^{+0.0061}_{-0.0061}$ & $0.0051^{+0.0058}_{-0.0058}$\\
\hline
$n_s$ & $0.962^{+0.014}_{-0.014}$ & $0.964^{+0.014}_{-0.014}$ \\
\hline
$\rm{ln}[10^{10}As]$ & $3.070^{+0.035}_{-0.035}$ &$3.066^{+0.035}_{-0.035}$ \\
\hline
$|D|$ & $ <0.711$(95\%{\rm CL}) & $ <0.131$(95\%{\rm CL})\\
\hline
\end{tabular}
\end{table*}

\section{conclusions\label{conclusions}}
In this work, we have studied the impact of the specific family of $f(R)$ models that can reproduce the same background expansion history as that of  $\Lambda$CDM model on the cosmic microwave background and the large scale structure using the full set of covariant cosmological perturbation equations. Based upon the covariant perturbation equations, we have modified the public available  Einstein-Boltzmann code CAMB and CosmoMC~\cite{mcmc} to conduct the Markov Chain Monte Carlo analysis on the parameter space and confront our $f(R)$ model with the latest observations. In the flat Universe, our results are in a good agreement with the previous independent work done within the parameterized framework of modified gravity. We have also extended our analysis to the non-flat Universe. From the fitting results, at the best fitted point we find that the curvature $\Omega_k$ term is slightly positive which is different from the results in the non-flat $\Lambda$CDM model\cite{WMAP}. Although more severe constraints on the $f(R)$ model can be obtained by adding the data
from cluster abundance, the halo mass function in our $f(R)$ model should be carefully studied before the data is used for our global fitting analysis. This will be an objective to our future work.

\emph{Acknowledgment: J.H.He would like to thank
B. R. Granett and L. Guzzo for helpful
discussions. J.H.He acknowledges the Financial
support of MIUR through PRIN 2008 and ASI through
contract Euclid-NIS I/039/10/0. }

\section{Appendix}
\subsection{The relation between $B_0$ and $D$}
Noting that
\begin{equation}
\frac{f_{RR}}{F}\frac{dR}{dx}\frac{H}{\frac{dH}{dx}}=\frac{\frac{\partial F}{\partial x}}{F}\frac{H}{\frac{\partial H}{\partial x}}\quad,
\end{equation}
using Eq.(\ref{Ex}) and Eq.(\ref{mG}) calculating straightforwardly, we obtain,
\begin{equation}
\begin{split}
B(x)&=\frac{2D p_+e^{3xp_+}(e^{3x}\Omega_d^0+\Omega_m^0)}{(\Omega_m^0)^2\left\{1+De^{3xp_+}{_2F_1}\left[q_+,p_+;r_+;-\frac{e^{3x}\Omega_d^0}{\Omega_m^0}\right]\right\}}\\
&\times\{\frac{q_+}{r_+}\Omega_d^0e^{3x}{_2F_1}\left[q_++1,p_++1;r_++1;-\frac{e^{3x}\Omega_d^0}{\Omega_m^0}\right]\\
&-\Omega_m^0{_2F_1}\left[q_+,p_+;r_+;-\frac{e^{3x}\Omega_d^0}{\Omega_m^0}\right]\}\quad.
\end{split}
\end{equation}
When $x=0$, we can find the explicit relationship between $D$ and $B_0$ as
\begin{equation}
\begin{split}
B_0&=\frac{2D p_+}{(\Omega_m^0)^2\left\{1+D{_2F_1}\left[q_+,p_+;r_+;-\frac{\Omega_d^0}{\Omega_m^0}\right]\right\}}\\
&\times\{\frac{q_+}{r_+}\Omega_d^0{_2F_1}\left[q_++1,p_++1;r_++1;-\frac{\Omega_d^0}{\Omega_m^0}\right]\\
&-\Omega_m^0{_2F_1}\left[q_+,p_+;r_+;-\frac{\Omega_d^0}{\Omega_m^0}\right]\}\quad.
\end{split}
\end{equation}

\subsection{Modifying CAMB}
Our work is based on the public available Einstein-Boltzmann code CAMB ~\cite{CAMB}. The basic equations in the CAMB code are based on the covariant approach in which it describes the cosmological perturbations in terms of the variables that are covariantly defined in the real universe(see \cite{cambequationsreviw} for reviews). When making $1+3$ decomposition of the physical quantities with respect to a family of observers, CAMB choose the observer coincide with the motion of CDM in the Universe, and the equations in CAMB are thus equivalent to the general perturbation equations as presented in~\cite{Kodama} to be fixed in the synchronous gauge. Here we summarize our cosmological perturbation
equations in the $f(R)$ gravity in the  synchronous gauge with the same conventions as that used in CAMB. In CAMB, the curvature perturbations are characterized by $\mathcal{Z}$ and $\sigma$
\begin{eqnarray}
\mathcal{Z}&=&\frac{h_L'}{2k}\nonumber\quad, \\
\sigma&=&k\alpha\nonumber \quad,
\end{eqnarray}
where
\begin{equation}
\eta_T'=\frac{k}{3}(\sigma-\mathcal{Z})\quad.
\end{equation}
The perturbed modified Einstein equations can be written as
\begin{eqnarray}
(F\mathcal{H}+\frac{1}{2}F')k\mathcal{Z}&=&\frac{\kappa^2}{2}a^2\delta \rho+Fk^2\eta_T\beta_2-\frac{3}{2}\mathcal{H}\delta F'\nonumber\\
&-&\frac{1}{2}\delta F k^2+\frac{3}{2}\mathcal{H}'\delta F\quad,\\
\frac{k^2}{3}F(\beta_2\sigma-\mathcal{Z})&=&\frac{\kappa^2}{2}a^2q+\frac{1}{2}k\delta F'\nonumber \\
&-&\frac{1}{2}k\mathcal{H}\delta F\quad,\\
\sigma'+2\mathcal{H}\sigma+\frac{F'}{F}\sigma&=&k\eta_T-\kappa^2a^2\frac{p\Pi}{F k}-k\frac{\delta F}{F}\quad,\\
\mathcal{Z}'+(\frac{1}{2}\frac{F'}{F}+\mathcal{H})\mathcal{Z}&=&(-k\beta_2+\frac{k}{2}+\frac{3\mathcal{H}^2}{k})\frac{\delta F}{F}\nonumber\\
     &-&\frac{\kappa^2a^2}{2kF}(\delta \rho+3\delta p)-\frac{3}{2}\frac{\delta F''}{kF}\quad,
\end{eqnarray}
where $\beta_2$ is the curvature factor
\begin{equation}
\beta_2=\frac{k^2-3K}{k^2}\quad.
\end{equation}
The propagation of the perturbed field $\delta F$ is given by
\begin{eqnarray}
&&\delta F''+2\mathcal{H}\delta F'+a^2(\frac{k^2}{a^2}+M^2)\delta F\nonumber\\
 &=&\frac{\kappa^2a^2}{3}(\delta \rho -3\delta p) -kF'\mathcal{Z}\quad,\label{deltF}
\end{eqnarray}
and the perturbation of the scalar curvature $\delta R$ is given by
\begin{equation}
a^2\delta R=2kZ'+6kZ\mathcal{H}-4k^2\beta_2\eta_T\quad.
\end{equation}

We have replaced the original perturbation equations with the above set of equations in the original CAMB code.
Another important part we need to modify is the source term of the CMB temperature anisotropy \cite{source}\cite{cambequations}
\begin{equation}
\begin{split}
&S_T(\tau,k)\\
&=e^{-\varepsilon}(\alpha''+\eta_T')\\
&+g(\Delta_{T0}+2\alpha'+\frac{v_b'}{k}+\frac{\zeta}{12\sqrt{\beta_2}}+\frac{\zeta''}{4k^2\sqrt{\beta_2}})\\
&+g'(\alpha+\frac{v_b}{k}+\frac{\zeta'}{2k^2\sqrt{\beta_2}})+\frac{1}{4}\frac{g''\zeta}{k^2\sqrt{\beta_2}}\\
&=e^{-\varepsilon}(\frac{\sigma''}{k}+\frac{k\sigma}{3}-\frac{k\mathcal{Z}}{3})\\
&+g(\Delta_{T0}+2\frac{\sigma'}{k}+\frac{v_b'}{k}+\frac{\zeta}{12\sqrt{\beta_2}}+\frac{\zeta''}{4k^2\sqrt{\beta_2}})\\
&+g'(\frac{\sigma}{k}+\frac{v_b}{k}+\frac{\zeta'}{2k^2\sqrt{\beta_2}})+\frac{1}{4}\frac{g''\zeta}{k^2\sqrt{\beta_2}}\quad,
\end{split}
\end{equation}
where $g=-\dot{\varepsilon}e^{-\varepsilon}=an_e\sigma_Te^{-\varepsilon}$ is the visibility function and $\varepsilon$ is the optical depth. $\zeta$ is given by
\begin{equation}
\zeta=(\frac{3}{4}I_2 +\frac{9}{2}E_2)\quad,
\end{equation}
where $I_2$, $E_2$ indicate the quadrupole of the photon intensity and the E-like polarization respectively ~\cite{cambequations}. In the original code, $\sigma$ and $\mathcal{Z}$ are calculated from the perturbation equations in the standard Einstein's gravity. In our work,  $\sigma$ and $\mathcal{Z}$ are calculated from perturbation equations in the $f(R)$ gravity.

\subsection{Cosmological perturbation in f(R) gravity}
The perturbed modified Einstein equations in the $f(R)$ gravity are given by
\begin{eqnarray}
-\frac{\kappa^2}{2}\delta \rho a^2&=&F[-k^2\phi+3\mathcal{H}(\mathcal{H}\psi-\phi')\nonumber\\
&+&k\mathcal{H}B+K(3\phi+\frac{1}{2}\mathcal{E}) -\frac{1}{6}k^2\mathcal{E}]\nonumber \\
    &+&F'(\frac{k}{2}B+3\mathcal{H}\psi-\frac{3}{2}\phi')-\frac{3}{2}\mathcal{H}\delta F'\nonumber \\
    &-&\delta F(\frac{3}{2}\mathcal{H}^2-\frac{3}{2}\frac{a''}{a}+\frac{k^2}{2})\quad.\label{JE_1}
\end{eqnarray}
\begin{eqnarray}
\kappa^2a^2\delta p&=&F[-\frac{2}{3}k^2\psi-\frac{2}{3}k^2\phi-\frac{1}{9}k^2\mathcal{E}\nonumber\\
&+&4\frac{a''}{a}\psi+\frac{2}{3}kB'+\frac{4}{3}k\mathcal{H}B-2\mathcal{H}^2\psi\nonumber\\
 &+&2\mathcal{H}\psi'-4\mathcal{H}\phi'-2\phi''+K(2\phi+\frac{1}{3}\mathcal{E})]\nonumber \\
 &+&F'[\frac{2}{3}kB+\psi'+2\mathcal{H}\psi-2\phi']\nonumber \\
 &+&\delta F[\mathcal{H}^2+\frac{a''}{a}-\frac{2}{3}k^2+2K]\nonumber \\
 &-&\delta F''-\delta F'\mathcal{H}+2\psi F''\quad.\label{JE_2}
\end{eqnarray}
\begin{eqnarray}
\kappa^2a^2p\Pi&=&F[-k^2\psi-k^2\phi+2k\mathcal{H}B+\mathcal{H}\mathcal{E}'\nonumber\\
&-&\frac{1}{6}k^2\mathcal{E}+\frac{1}{2}\mathcal{E}''+kB']\nonumber \\
&+&F'(\frac{1}{2}\mathcal{E}'+kB)-k^2\delta F\quad.\label{JE_3}
\end{eqnarray}
\begin{eqnarray}
-\frac{1}{2}\kappa^2a^2q&=&F[k\phi'-k\mathcal{H}\psi+2\mathcal{H}^2B\nonumber\\
&-&\frac{a''}{a}B+\frac{1}{6}k\mathcal{E}'-\frac{1}{2}\mathcal{E}'\frac{K}{k}]\nonumber \\
&+&F'[\mathcal{H}B-\frac{1}{2}k\psi]+\frac{1}{2}k\delta F'\nonumber \\
&-&\frac{1}{2}k\mathcal{H}\delta F-\frac{1}{2}B F''\quad,\label{JE_4}
\end{eqnarray}
where $q=(\rho+p)v$. The above perturbation equations are self-consistent covariant equations.  We can show that under the infinitesimal coordinate transformation, namely, Eq.(\ref{transformation}),
the perturbation equations could keep the same form.
\begin{eqnarray}
\hat{\psi}Y^{(s)}&=& (\psi-{\xi^0}'-\mathcal{H}\xi^0)Y^{(s)}\nonumber\quad,\\
\hat{\phi}Y^{(s)}&=& (\phi-\frac{1}{3}k\beta-\mathcal{H}\xi^0)Y^{(s)}\nonumber \quad,\\
\hat{B}Y^{(s)}_i&=&(B-k\xi^0-\beta')Y^{(s)}_i\nonumber\quad, \\
\hat{\mathcal{E}}Y^{(s)}_{ij}&=&(\mathcal{E}+2k\beta)Y^{(s)}_{ij}\quad.\label{transformation}
\end{eqnarray}
The perturbation for the scalar field $\delta F$ satisfies
\begin{equation}
\begin{split}
\delta F&=f_{RR}\delta R \quad, \\
f_{RR}&=\frac{\partial F}{\partial R}\quad,\nonumber
\end{split}
\end{equation}
where the perturbation of scalar curvature $\delta R$ is given by
\begin{equation}
\begin{split}
a^2\delta R&=-12\psi\frac{a''}{a}-6\mathcal{H}\psi'+4\phi k^2-2K\mathcal{E} \\
&+2\psi k^2+18\mathcal{H}\phi'+\frac{2}{3}\mathcal{E}k^2+6\phi''\\
&-6B\mathcal{H}k-12\phi K-2B'k\quad .\label{deltaR}
\end{split}
\end{equation}

Inserting Eq.(\ref{deltaR}) into Eq.(\ref{JE_2}) to eliminate $\phi''$ and using Eq.(\ref{JE_1}) to eliminate
$k^2\phi$, the equation governing the behavior of $\delta F$ gives rise to
\begin{equation}
\begin{split}
\delta F''&+2\mathcal{H}\delta F'+a^2(\frac{k^2}{a^2}+M^2)\delta F\\
&=\frac{\kappa^2a^2}{3}(\delta \rho-3\delta p) +2\psi F'' \\
&+F'(4\mathcal{H}\psi-3\phi'+\psi'+kB)\quad,\label{gdeltF}
\end{split}
\end{equation}
where
\begin{equation}
M^2=\frac{1}{3}(\frac{F}{f_{RR}}-R)\label{Mdef}\quad,
\end{equation}
and
\begin{equation}
R=\frac{6a''}{a^3}+\frac{6K}{a^2}\quad.
\end{equation}

\end{document}